\documentclass[journal]{IEEEtran}
\ifCLASSINFOpdf
\else
\fi
\usepackage{cite}
\usepackage[draft]{hyperref}
\usepackage{amsmath}
\usepackage{amsfonts}
\usepackage{graphicx}
\usepackage{multirow}
\usepackage{float}
\usepackage{wrapfig}
\usepackage{hyperref}
\usepackage{graphicx}
\usepackage{caption}
\usepackage{subcaption}
\usepackage{epstopdf}
\usepackage{color}
\usepackage{longtable}
\usepackage{moreverb}
\usepackage{amsmath}
\usepackage{algorithmicx}
\usepackage{algorithm}
\usepackage{algpseudocode}
\usepackage[labelsep=period]{caption}
\usepackage{textcase}
\renewcommand{\figurename}{Fig.}
\renewcommand\IEEEkeywordsname{Index Terms}
\algdef{SE}[DOWHILE]{Do}{doWhile}{\algorithmicdo}[1]{\algorithmicwhile\ #1}%
\usepackage{listings}
\usepackage{esvect}

\begin{document}
%
\title{Exploiting LTE-Advanced HetNets and FeICIC for UAV-assisted Public Safety Communications}
%
%
%

\author{\IEEEauthorblockN{Abhaykumar Kumbhar$^{1,2}$, \.{I}smail~G\"uven\c{c}$^{3}$, Simran Singh$^{3}$, and Adem~Tuncer$^{4}$}\\
\IEEEauthorblockA{$^1$Dept. Electrical and Computer Engineering, Florida International University, Miami, FL, 33174\\
$^2$Motorola Solutions, Inc., Plantation, FL, 33322\\
$^3$Dept. Electrical and Computer Engineering, North Carolina State University, Raleigh, NC, 27606\\
$^4$Yalova University, \c{C}{\i}narc{\i}k Yolu \"Uzeri, 77200 Merkez/Yalova, Turkey\\
\vspace{-0.5cm}
}}

\maketitle

\begin{abstract}
Ensuring ubiquitous mission-critical public safety communications (PSC) to all the first responders in the public safety network (PSN) is crucial at an emergency site. Recently, the use of unmanned aerial vehicles (UAVs) has received extensive interest for PSC to fill the coverage holes and establish reliable connectivity. The UAVs can be deployed as unmanned aerial base stations (UABSs) as part of a heterogeneous network (HetNet) PSC infrastructure. In this article, we design a PSC LTE-Advanced HetNet for different path loss models and deployment mechanism for UABSs. We enhance the system-wide spectral efficiency (SE) of this PSC HetNet by apply cell range expansion (CRE) to UABSs and mitigating the inter-cell interference arising in the HetNet by applying 3GPP Release-10 enhanced inter-cell interference coordination (eICIC) and 3GPP Release-11 further-enhanced inter-cell interference coordination (FeICIC). Through Monte-Carlo simulations, we compare the system-wide 5th percentile SE when UABSs are deployed on a hexagonal grid and when their locations are optimized using a genetic algorithm, while also jointly optimizing the CRE and the inter-cell interference coordination parameters. Our results show that at optimized UABS locations, reduced power subframes (FeICIC) defined in 3GPP Release-11 can provide considerably better 5th percentile SE than the 3GPP Release-10 with almost blank subframes (eICIC).
\end{abstract}


%
\IEEEpeerreviewmaketitle

\section{Introduction}
Public safety communications (PSC) is considered to be the cornerstone of public safety response system and plays a critical role in saving lives, property, and national infrastructure during a natural or man-made emergency. To further enhance the capabilities of next generation PSC networks, countries such as the United States, the United Kingdom, and Canada are building a 4G Long Term Evolution (LTE) based broadband public safety network~\cite{R1,athukoralage2016regret,UKESNCurrentStatus}. In particular, 4G mobile networks have great potential to revolutionize PSC during emergency situations by providing much needed high-speed real-time data, video, and multimedia services along with mission-critical communication. However, given the limited spectrum allocation for LTE-based PSC, the usage of such services would need improved channel capacity quality, and spectral efficiency (SE) \cite{yuksel2016pervasive,R1}.

An additional challenge in designing an LTE-based PSC network is to achieve seamless and ubiquitous coverage by minimizing coverage holes. However, this coverage criteria will be difficult to achieve using only the dedicated cell towers. For example, the first responders network authority (FirstNet) in the United States is designing a PSC network, which is required to minimize coverage gaps and attain at least $95\%$ coverage of the geographical area and human population~\cite{federaldocket}.

\begin{figure} [t]
\centering
\includegraphics[width=0.85\linewidth]{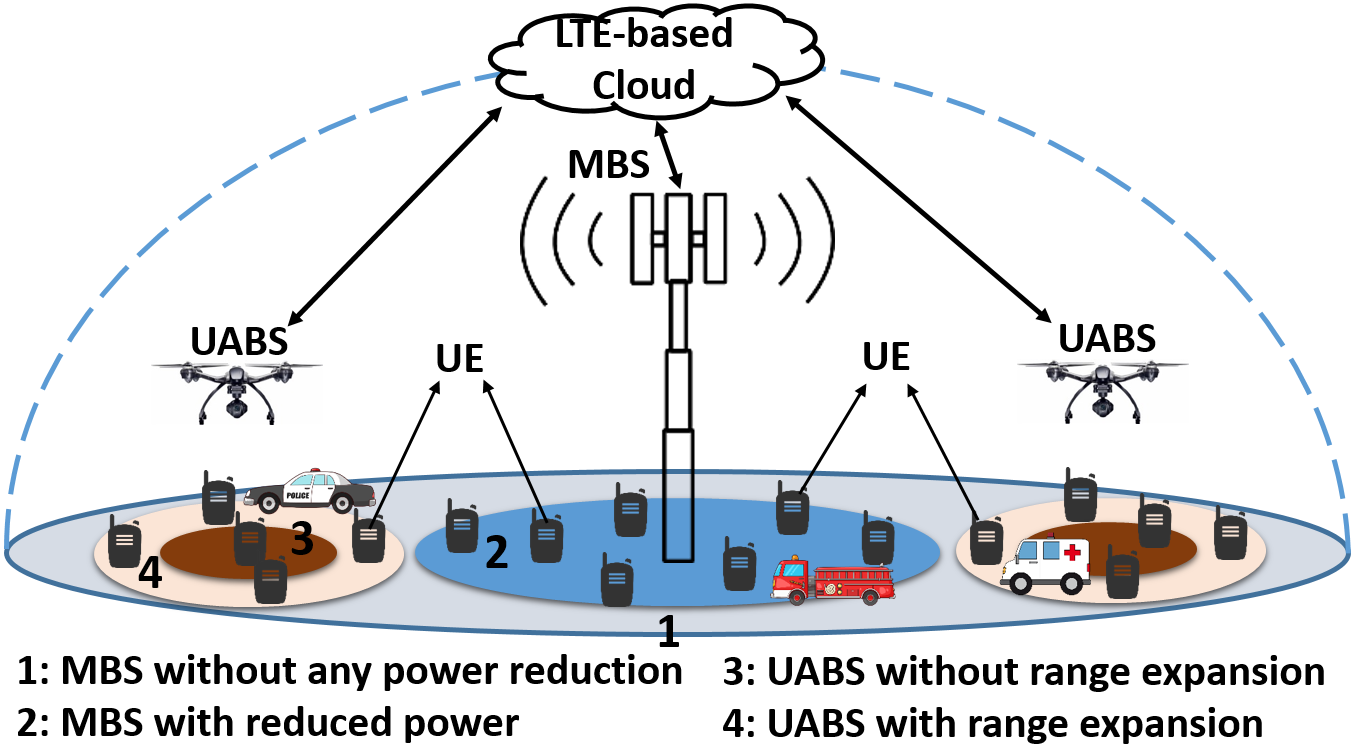}
\caption{The PSC scenario with MBS and UABSs constituting an air/ground HetNet infrastructure. The MBS can use inter-cell interference coordination techniques defined in LTE-Advanced. The UABSs can dynamically change their position to maintain good coverage and can utilize range expansion bias to take over MBS UEs.}
\label{PscHetnet}\vspace{-0.45cm}
\end{figure}

To this end, the deployment of LTE-Advanced small cells is increasingly becoming popular to provide improved spectral capacity and extend network coverage~\cite{VZWSmallCell,ATTSmallCell,nakamura2013trends,merwaday2016improved,R10}. Moreover, recent practices in small cell deployment include the usage of ad-hoc and air-borne small cells such as the unmanned aerial base stations (UABSs). With minimum interdependencies and at low cost, these UABSs can be deployed at an emergency site for restoration and temporary expansion of PSC network~\cite{R4}. This can improve overall network spectral capacity and provides virtually omnipresent coverage which is essential for first-responders to be efficient and save lives. On the other hand, the deployment of these UABSs can also introduce significant inter-cell interference with the ground network and also limit the overall PSC network SE~\cite{moore2014first, merwaday2016improved, R10, shifat2017game}. 

In this paper, for a suburban environment, we study the performance of inter-cell interference coordination (ICIC) and cell range expansion (CRE) techniques for an LTE band class~14 PSC network~\cite{R1} as shown in Fig.~\ref{PscHetnet}. By randomly removing macro base stations (MBSs), we simulate a mock emergency situation to study the impact of interference and CRE when the UABSs are deployed. The simulation model considers  Rayleigh fading and simplified path loss model and Okumura-Hata model for calculating the propagation losses. Subsequently, we explore potential gains in 5th percentile SE (5pSE) from the use of Release-10/11 ICIC techniques for a UABS-based PSC HetNet. Furthermore, we jointly optimize the UABSs locations,  CRE parameter of the UABSs,  and ICIC parameters for both MBSs and UABSs.

The rest of this paper is organized as follows. Section~\ref{Sec:LitReview} provides a brief literature review of related work, Section~\ref{systemModel} introduces the UABS-based HetNet model, different path-loss models, and the definition of 5pSE as a function of network parameters. The UABSs deployment and ICIC parameter configurations using the genetic algorithm and hexagonal grid UABS model are described in Section~\ref{sec:UabsDeploy}. In Section~\ref{simulation}, we analyze and compare the 5pSE of the HetNet using extensive computer simulations for various ICIC techniques, and finally, the last section provides some concluding remarks. Table~\ref{tab:NotSymb} lists the notations and symbols used  throughout the paper.  

\section{Literature review}\label{Sec:LitReview}
LTE based HetNets are a well-researched topic, and numerous studies have been carried to enhance the network performance, improve coverage, and mitigate network interference. In ~\cite{saquib2013fractional}, a fractional frequency reuse method is used to mitigate interference in a fixed HetNet, improve the indoor coverage, and maximize the network SE by minimizing the UE's outage probability to cell-edge UEs in an OFDMA-based LTE HetNet. However,~\cite{saquib2013fractional} did not consider any of the 3GPP Release-10 and Release-11 inter-cell interference coordination (ICIC) techniques for fixed HetNet deployments. 

\begin{table}[t]
\caption{Notations and symbols used in system model.}
\label{tab:NotSymb}
\centering
\small
\begin{tabular}{p{2cm} p{6cm}}
\hline
Symbol & Description\\
\hline
$\rm \lambda_{\rm mbs}$, $\rm \lambda_{\rm ue}$   & Intensities of the MBS and UE nodes\\
${ \rm {{\bf X}_{mbs}}}, {\bf {X_{\rm ue}}}$      & Locations of MBS and UE.\\
$P_{\rm mbs},P_{\rm uabs}$                        & Maximum transmit power of MBS and UABS\\
$P^\prime_{\rm mbs},P^\prime_{\rm uabs}$          & Effective transmit power of MBS and UABS\\
$K, K^\prime$				                      & Attenuation factors due to geometrical parameters of antennas for both MBS and UABS\\
$H$                                               & Exponentially distributed random variables that account for Rayleigh fading \\
$\delta$                                          & Path loss exponent (PLE) \\
$f_{\rm c}$                                       & Carrier frequency (LTE band class 14) \\
$h_{\rm bs}$                                      & Height of the base station in Okumura-Hata model\\
$h_{\rm ue}$                                      & Height of a UE in Okumura-Hata model\\ 
$d_{mn}, d_{mu}$                                  & Distance of a UE from its MOI and UOI, respectively \\
$S_{\rm mbs}(d_{mn})$                             & RSRP from the MOI \\
$S_{\rm uabs}(d_{mu})$                            & RSRP from the UOI\\
$Z$                                               & Total interference at a UE from USF and CSF, respectively \\
$\gamma, \gamma^\prime$                           & SIR from MOI and UOI, respectively during USF \\
$\gamma_{\rm csf},\gamma^\prime_{\rm csf}$        & SIR from MOI and UOI, respectively during CSF \\  
$\alpha$                                          & Power reduction factor for MBS during the transmission of CSFs \\
$\beta$                                           & Duty cycle for the transmission of USF\\
$\tau$                                            & Cell range expansion bias\\
$\rho, \rho\prime$                                & Scheduling threshold for MUE and UUE, respectively\\
$N_{\rm usf}^{\rm mbs}, N_{\rm csf}^{\rm mbs}$    & Number of USF-MUEs and CSF-MUEs, respectively in a cell\\
$N_{\rm usf}^{\rm uabs}, N_{\rm csf}^{\rm uabs}$  & Number of USF-UUEs and CSF-UUEs, respectively in a cell\\
$C_{\rm usf}^{\rm mbs}, C_{\rm csf}^{\rm mbs}$    & Aggregate SEs for USF-MUEs and CSF-MUEs, respectively in a cell\\
$C_{\rm usf}^{\rm uabs}, C_{\rm csf}^{\rm uabs}$  & Aggregate SEs for USF-UUEs and CSF-UUEs, respectively in a cell\\
$\hat{\bf X}_{\rm uabs}^{(\rm hex)}$              & Fixed hexagonal locations of deployed UABS\\
$\hat{\bf X}_{\rm uabs}$                          & Optimized UABS locations computed using GA\\
${\bf S}^{\rm ICIC}_{\rm mbs}$                              & Matrix of ICIC parameters for MBSs\\
${\bf S}^{\rm ICIC}_{\rm uabs}$                          & Matrix of ICIC parameters for UABSs\\
\hline
\end{tabular}
\end{table}

3GPP Release-10 eICIC and Release-11 FeICIC techniques have been studied in \cite{deb2014algorithms,R10,mukherjee2011effects} for HetNets. For example,~\cite{deb2014algorithms} proposes algorithms that jointly optimizes the eICIC parameters, UE cell association rules, and the spectrum resources shared between the macro and fixed small cells. Nevertheless, 3GPP Release-11 FeICIC technique for better radio resource utilization and CRE for offloading a larger number of UEs to small cells was not considered in~\cite{deb2014algorithms}. The effectiveness of 3GPP Release-10 and Release-11 ICIC techniques with ICIC parameter optimization have been studied in~\cite{R10}, without considering any mobility for small cells. 

Recent advancements in UAV technology has enabled the possibility of deploying small cells as UABSs mounted with a communication system. UABSs such as balloons, quadcopters, and gliders equipped with LTE-Advanced capabilities can be utilized to further enhance the capabilities of LTE-based HetNets. The ability of UABSs to dynamically reposition in a HetNet environment can improve the overall SE of the network by filling the coverage gaps and offloading UEs in high-traffic regions. Hence, it is of critical nature to optimize the locations of UABSs in a UAV-based HetNet to optimize SE gains. 

Recent studies~\cite{al2014optimal,bor2016efficient,mozaffari2016optimal,sharma2016uav,naranguav,christy2017optimum} 
are mainly focused on finding an optimal location of UAVs in the geographical area of interest to meet traffic demands. In~\cite{al2014optimal,bor2016efficient,mozaffari2016optimal,sharma2016uav}, UAV location optimization have been explored; however, inter-cell interference coordination techniques are not explicitly taken into account. Authors in~\cite{merwaday2015uav, merwaday2016improved} explore UABS-assisted LTE HetNets, where  the UABSs use CRE for offloading users from a macrocell but do not consider any ICIC in the cell expanded region. To maximize the 5pSE of the HetNet, a brute force method is used to find the optimal UAV locations in~\cite{merwaday2015uav}, while the genetic algorithm is used for optimizing UAV locations in ~\cite{merwaday2016improved}.

The effect of interference in a UAV-based network is investigated in~\cite{mozaffari2015drone}. By calculating the optimal distance between the two interfering UAVs, each UAV is positioned at fixed height to maximize the coverage area. However, this UAV-based network is not designed for LTE-Advanced HetNets. A priority-based UE offloading and UE association with mobile small cells for PSC is studied in~~\cite{kaleem2016public}. To improve the overall system throughput, 3GPP Release-10 eICIC and CRE is taken into account. However, using almost blank subframes (ABS) at an MBS results in under-utilization of radio resources when compared to the use of reduced power FeICIC.

To the best of our knowledge, the use of 3GPP Release-10/11 inter-cell interference coordination techniques in UAV-based LTE-Advanced HetNets has not been adequately studied in the literature. The contribution of our study is to address this challenge of interference mitigation by using and optimizing FeICIC and CRE techniques and at the same time, optimize the UAV locations using the genetic algorithm to achieve maximum SE.

\begin{figure}[t]
\centering
\begin{subfigure}[b]{0.38\textwidth}
\label{undestroyedInfra}
\includegraphics[width=1\textwidth]{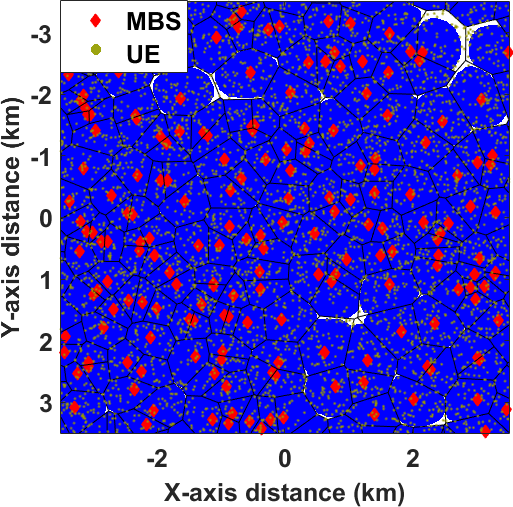}
\caption{Typical PSC network.}
\end{subfigure}
\begin{subfigure}[b]{0.38\textwidth}
\label{destroyedInfra}
\includegraphics[width=1\textwidth]{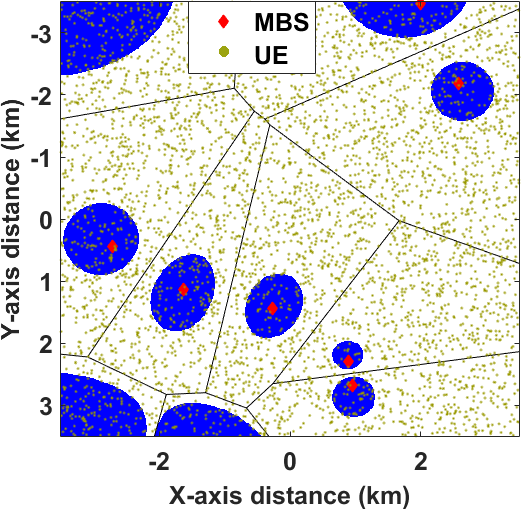}
\caption{PSC network after a disaster.}
\end{subfigure}
\caption{Wireless network SE coverage before/after a disaster.}
\label{fig:infrastructure1}
\vspace{-2mm}
\end{figure}

\section{System Model}
\label{systemModel}

In this paper, we consider a wireless network after a disaster as shown in Fig.~\ref{fig:infrastructure1}. In particular, Fig.~\ref{fig:infrastructure1}(a) shows that most of the geographical area in a  typical PSC network is under SE coverage before a disaster. In the event of a disaster, the PSC network infrastructure is destroyed, and the first responders and victim users experience SE outage as illustrated by white areas in Fig.~\ref{fig:infrastructure1}(b). In this scenario, the existing MBSs get overloaded with many UEs, and as a result, these UEs begin to experience poor QoS. Subsequently, at the site of emergency, the first responders and victim users located in the SE outage regions will observe very low SE or possibly complete outage. 

To address the SE problems in a scenario as in Fig.~\ref{fig:infrastructure1}(b), We consider a two-tier HetNet deployment with MBSs and UABSs as shown in Fig.~\ref{PscHetnet}, where all the MBS and UABS locations (in three dimensions) are captured in matrices ${\bf X}_{\rm mbs} \in \mathbb{R}^{N_{\rm mbs}\times 3}$  and ${\bf X}_{\rm uabs}\in \mathbb{R}^{N_{\rm uabs}\times 3}$, respectively, where $N_{\rm mbs}$ and $N_{\rm uabs}$ denote the number of MBSs and UABSs within the simulation area. The MBS and user equipment (UE) locations are each modeled using a two-dimensional Poisson point process (PPP) with intensities $\rm \lambda_{\rm mbs}$ and $\rm \lambda_{\rm ue}$, respectively~\cite{R10, R12}. The UABSs are deployed at fixed height and their locations are either optimized using the genetic algorithm, or they are deployed on a fixed hexagonal grid. The heights for each of these wireless nodes are specified in Table~\ref{tab:SysParams}.

We assume that the MBSs and the UABSs share a common transmission bandwidth, round robin scheduling is used in all downlink transmissions, and full buffer traffic is used in every cell. The transmit power of the MBS and UABS are $P_{\rm mbs}$ and $P_{\rm uabs}$, respectively, while $K$ and $K^\prime$ are the attenuation factors due to geometrical parameters of antennas for the MBS and the UABS, respectively. Then, the effective transmit power of the MBS is $P^\prime_{\rm mbs} = KP_{\rm mbs}$, while the effective transmit power of the UABS is $P^\prime_{\rm uabs} = K^\prime P_{\rm uabs}$.

An arbitrary UE $n$ is always assumed to connect to the nearest MBS or UABS, where $n\in\{1,2,...,N_{\rm ue}\}$. Then, for the $n$th UE the reference symbol received power (RSRP) from the macro-cell of interest (MOI) and the UAV-cell of interest (UOI) are given by~\cite{R10}
\begin{align}
     S_{\rm mbs}(d_{mn}) = \frac{P^\prime_{\rm mbs}H}{10^{\varphi/10}}, \ S_{\rm uabs}(d_{un}) = \frac{P^\prime_{\rm uabs}H}{10^{\varphi^\prime/10}},
\end{align}
where the random variable $H\sim$ Exp(1) accounts for Rayleigh fading, $\varphi$ is the path-loss observed from MBS in dB, $\varphi^\prime$ is the path-loss observed from UABS in dB, $d_{mn}$ is the distance from the nearest MOI, and $d_{un}$ is the distance from the nearest UOI. In this article, the variation in deployment height of the UABSs and the effects of dominant line-of-sight links on 5pSE of the network are not treated explicitly and left as a future work. Hence, Rayleigh fading channel is considered instead of Rican fading channel in this system model.

\subsection{Path Loss Model}
Given the complexity of signal propagation, an accurate path-loss modeling would require complex ray tracing models and empirical measurements. 
To measure the path-loss observed by the $n$th UE, we use the simplified path loss model (SPLM), which is an approximation to the real propagation channel. However, for more accurate analysis of the signal reliability requirements in PSC networks, we also implement the free space suburban Okumura-Hata path loss model (OHPLM) with LTE band class 14 frequency~\cite{R1,newman2010fcc}.

\subsubsection{Simplified Path Loss Model}
The SPLM gives a coarse analysis of signal propagation and is a function of path loss exponent and the distance between the serving base station and the  $n$th UE ~\cite{tsai2011path}. The SPLM is a free space model and does not considered any physical structures or other obstacles which might affect the coverage of UABSs in real world deployments. Based on the SPLM, the path-loss (in dB) observed by the $n$th UE from $m$th MOI and $u$th UOI is given by
\begin{align}
\varphi = 10{\rm log}_{10}(d_{mn}^\delta), \ {\varphi^\prime} = 10{\rm log}_{10}(d_{un}^\delta) \label{Eq:SPLM},
\end{align}
where $\delta$ is the path-loss exponent, and $d_{un}$ depends on the locations of the UABSs that will be dynamically optimized. 

\begin{figure} [!htbp]
\centering
\begin{subfigure}[b]{0.32\textwidth}
\label{undestroyedInfra}
\includegraphics[width=1.04\textwidth]{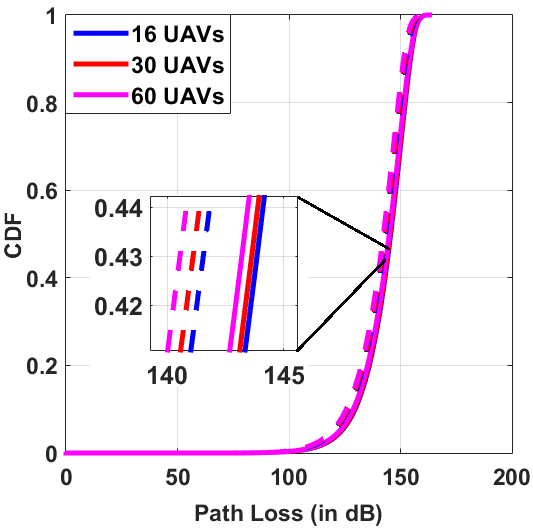}
\centering
\caption{CDF of simplified path loss model.}
\end{subfigure}
\begin{subfigure}[b]{0.33\textwidth}
\label{destroyedInfra}
\includegraphics[width=1.05\textwidth]{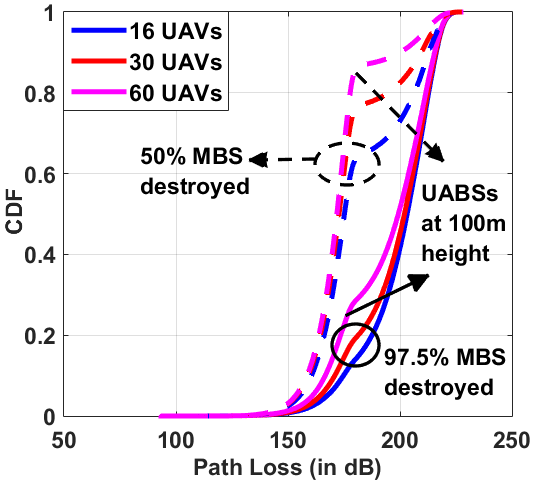}
\centering
\caption{CDF of Okumura-Hata model.}
\end{subfigure}
\caption{The CDF of the combined path loss observed from all the base stations. Dashed lines correspond to the scenario with $50\%$ of the MBS destroyed, while solid lines correspond to the scenario with $97.5\%$ of the MBS destroyed.}
\label{fig:pathloss}
\end{figure}

Fig.~\ref{fig:pathloss}(a) shows the empirical path-loss cumulative distribution functions (CDFs), calculated for all the distances between base stations (${\bf X}_{\rm mbs}$ and ${\bf X}_{\rm uabs}$) and UEs (${\bf X}_{\rm ue}$) using ~\eqref{Eq:SPLM}. For the path-loss exponent value defined in Table~\ref{tab:SysParams}, we plot the CDFs for the cases when $50\%$  and $97.5\%$ of the MBS are destroyed. Inspection of Fig.~\ref{fig:pathloss}(a) reveals the variation in CDFs is minimum for the different number of UABSs deployed and for a different number of the MBSs destroyed. This is because the SPLM does not consider external terrestrial factors. Nevertheless, the maximum allowable path-loss for the system is 160 dB.

\subsubsection{Okumura-Hata Path Loss Model}
\label{ohplm}
The OHPLM is more suited for a terrestrial environment with man-made structures and the environment in which the base-station height does not vary significantly~\cite{xiroOnline,ranvier2004path}. This model is a function of the carrier frequency, distance between the UE and the serving cell, base station height, and UE antenna height~\cite{xiroOnline,mollel2014comparison}. Based on curve fitting of Okumura's original results, the path-loss (in dB) observed by the $n$th UE from MOI and UOI is given by~\cite{molisch2012wireless,alqudah2016validation} 
\begin{align}
\varphi =  A + B{\rm log}(d_{mn}) + C, \label{Eq:PL}\\
\varphi^\prime = A + B{\rm log}(d_{un}) + C \label{Eq:PLprime},
\end{align}
where the distances $d_{mn}$ and $d_{un}$ are in km, and the factors $A$,$B$, and $C$ depend on the carrier frequency and antenna height. 

In a suburban environment, the factors $A$, $B$, $C$ are given~by 
\begin{align}
&     A = 69.55 + 26.16{\rm log}(f_{\rm c}) - 13.82{\rm log}(h_{\rm bs}) - a(h_{\rm ue}), \label{Eq:A} \\
&     B = 44.9 - 6.55{\rm log}(h_{\rm bs}), \label{Eq:B}\\
&     C = -2{\rm log}(f_{\rm c}/28)^2 - 5.4 \label{Eq:C},
\end{align}
where $f_{\rm c}$ is the carrier frequency in MHz, $h_{\rm bs}$ is the height of the base station in meter, and $a(h_{\rm ue})$ is the correction factor for the UE antenna height $h_{\rm ue}$ in meter, which is defined as  
\begin{align}
&    a(h_{\rm ue}) = 1.1{\rm log}(f_{\rm c}) - 0.7h_{\rm ue} - 1.56{\rm log}(f_{\rm c}) - 0.8~. \label{Eq:ahue}
\end{align}

Furthermore, OHPLM assumes the carrier frequency ($f_{\rm c}$) to be between 150 MHz to 1500 MHz, the height of the base station ($h_{\rm bs}$) between 30 m to 200 m, UE antenna height ($h_{\rm ue}$) between 1 m to 10 m, and the distances $d_{mn}$ and $d_{un}$ between 1 km to 10 km~\cite{molisch2012wireless,alqudah2016validation}. Nevertheless, the simulation values considered for OHPLM are defined in Table~\ref{tab:SysParams}. 

In Fig.~\ref{fig:pathloss}(b), we plot the empirical path-loss CDFs using \eqref{Eq:PL}--\eqref{Eq:ahue} and the OHPLM parameters in Table~\ref{tab:SysParams}. Moreover, we plot the OHPLM path-loss CDFs for the cases when $50\%$  and $97.5\%$ of the MBS are destroyed. Inspection of Fig.~\ref{fig:pathloss}(b) reveals a step-wise distribution of path-loss in the CDFs. This behavior is due to the variation in the height of base-stations, i.e., UABSs are deployed at the height of 100~m (larger path loss) while the height of MBSs is 30~m (smaller path loss). With $50\%$ of the MBS destroyed, it can be seen that most UEs are connected to the MBSs, while with $97.5\%$ of the MBS destroyed most UEs are served by the UABSs.  Regardless, the maximum allowable path-loss when 50\% and 97.5\% of the MBSs are destroyed is 225 dB as shown in Fig.~\ref{fig:pathloss}(b).

\subsection{3GPP Release-10/11 Inter-Cell Interference Coordination}
\label{icicidetails}
Due to their low transmission power, the UABSs are unable to associate a larger number of UEs compared to that of MBSs. However, by using the cell range expansion (CRE) technique defined in 3GPP Release-8, UABSs can associate a large number of UEs by offloading traffic from the MBSs. A negative side effect of CRE includes increased interference in the downlink of cell-edge UEs or the UEs in CRE region of the UABS, which is addressed by using ICIC techniques in LTE and LTE-Advanced~\cite{R5, R7, R8}.

\begin{figure} [t]
\begin{subfigure}[b]{1\linewidth}
\centering
\includegraphics[width=0.95\linewidth]{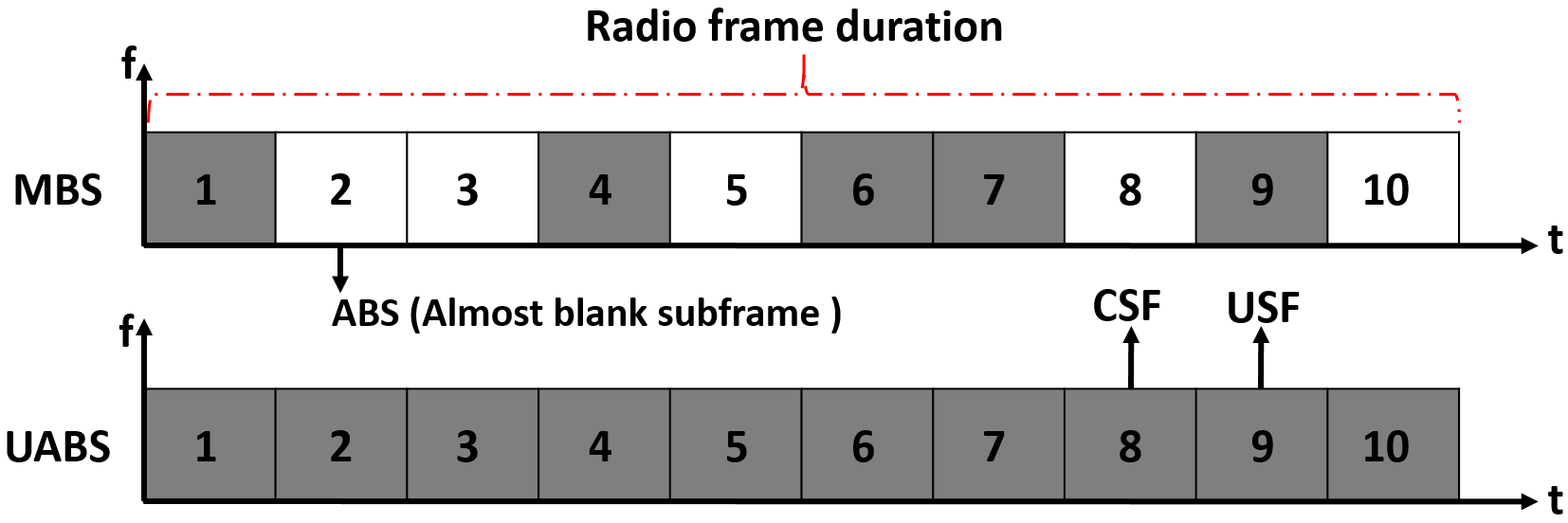}
\caption{3GPP Release-10 eICIC with ABS.}
\label{ReducedPowerFrames}
\end{subfigure}
\begin{subfigure}[b]{1\linewidth}
\centering
\includegraphics[width=0.95\linewidth]{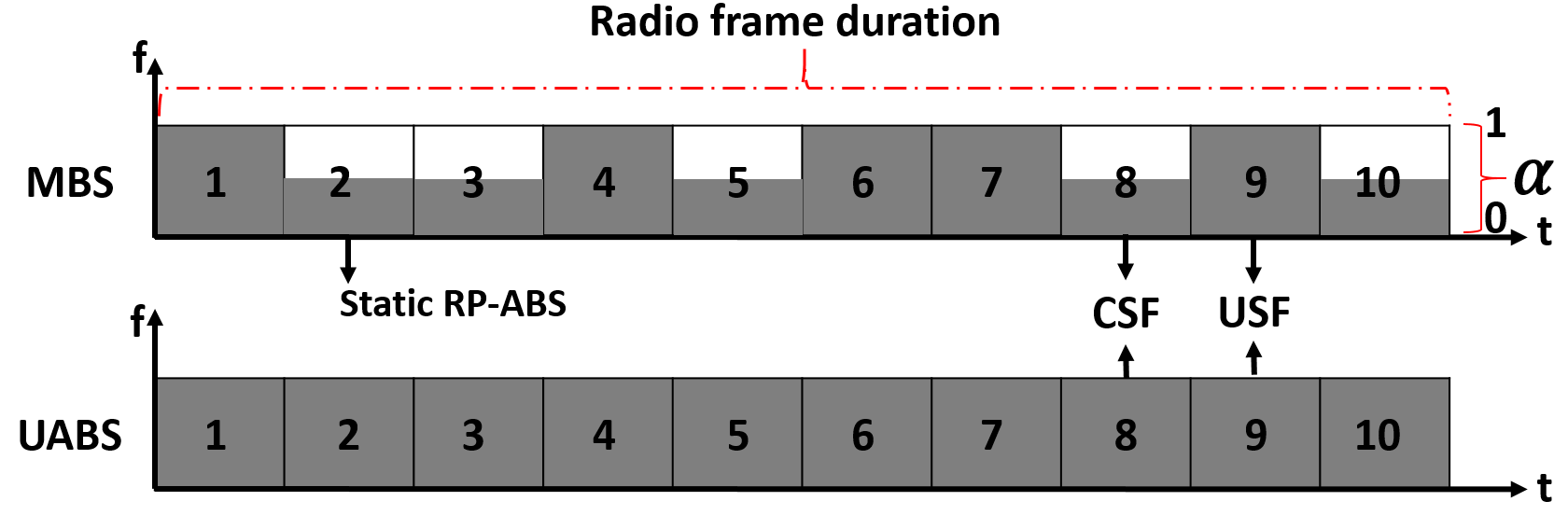}
\caption{3GPP Release-11 FeICIC with reduced power ABS (RP-ABS).}
\label{StaticPrAbs}
\end{subfigure}
\caption{LTE-Advanced frame structures for time-domain ICIC.}\label{Fig2} \vspace{-0.25cm}
\end{figure}

3GPP Release-10 introduced a time-domain based enhanced ICIC (eICIC) technique to address interference problems. In particular, it uses ABSs which require the MBS to completely blank the transmit power on the physical downlink shared channel (PDSCH) resource elements as shown in Fig.~\ref{Fig2}(a). This separates the radio frames into coordinated subframes (CSF) and uncoordinated subframes (USF). On the other hand, 3GPP Release-11 defines further-enhanced ICIC (FeICIC), where the data on PDSCH is still transmitted but at a reduced power level as shown in Fig.~\ref{Fig2}(b).

The MBSs can schedule their UEs either in USF or in CSF based on the scheduling threshold $\rho$. Similarly, the UABSs can schedule their UEs either in USF or in CSF based on the scheduling threshold $\rho^\prime$.
Let $\beta$ denote the USF duty cycle, defined as the ratio of number of USF subframes to the total number of subframes in a radio frame. Then, the duty cycle of CSFs is $(1-\beta)$. For ease of simulation, in this paper, we consider that the USF duty cycle $\beta$ is fixed at $0.5$ for all the MBSs, which is shown in~\cite{R10} to have limited effect on system performance when $\rho$ and $\rho'$ are optimized. Finally, let $0\leq\alpha\leq 1$ denote the power reduction factor in coordinated subframes of the MBS for the FeICIC technique; $\alpha=0$ corresponds to Release-10 eICIC, while $\alpha=1$ corresponds to no ICIC (e.g., as in LTE Release-8). We assume that the ABS and reduced power pattern are shared via the X2 interface, which is a logical interface between the base stations. Implementation of the X2 interface for UABSs is left as a future consideration.

Given the eICIC and FeICIC framework in 3GPP LTE-Advanced as in Fig.~\ref{Fig2}, and following an approach similar to that in~\cite{R10} for a HetNet scenario, the signal-to-interference ratio (SIR) experienced by an arbitrary UE can be defined for CSFs and USFs for the MOI and the UOI as follows:
\begin{align}
\Gamma &= \frac{S_{\rm mbs}(d_{mn})}{S_{\rm uabs}(d_{un}) + Z},\rightarrow {\rm USF\ SIR\ from\ MOI} \label{Eq:SIR1} \\
\Gamma_{\rm csf} &= \frac{\alpha S_{\rm mbs}(d_{mn})}{S_{\rm uabs}(d_{un}) + Z} \rightarrow {\rm CSF\ SIR\ from\ MOI}, \label{Eq:SIR2} \\
\Gamma^\prime &= \frac{S_{\rm uabs}(d_{un})}{ S_{\rm mbs}(d_{mn}) + Z} \rightarrow {\rm USF\ SIR\ from\ UOI}, \label{Eq:SIR3} \\
\Gamma^\prime_{\rm csf} &= \frac{S_{\rm uabs}(d_{un})}{\alpha S_{\rm mbs}(d_{mn})+ Z} \rightarrow {\rm CSF\ SIR\ from\ UOI},\label{Eq:SIR4}
\end{align}
where $Z$ is the total interference power at a UE during USF or CSF from all the MBSs and UABSs, excluding the MOI and the UOI. In hexagonal grid UABS deployment model (and in~\cite{R10}), locations of the UABSs (and small cells) are fixed. To maximize the 5pSE of the network, in this paper we actively consider the SIRs in~\eqref{Eq:SIR1}--\eqref{Eq:SIR4} while optimizing the locations of the UABSs using the genetic algorithm.

\subsection{UE Association and Scheduling}
The cell selection process relies on $\Gamma$ and $\Gamma^\prime$ in \eqref{Eq:SIR1} and \eqref{Eq:SIR3}, respectively, for the MOI and UOI SIRs, as well as the CRE $\tau$. If $\tau\rm\Gamma^\prime$ is less than $\rm \Gamma$, then the UE is associated with the MOI; otherwise, it is associated with the UOI. After cell selection, the MBS-UE (MUE) and UABS-UE (UUE) can be scheduled either in USF or in CSF radio subframes as: 
\begin{align}
& \rm If\ \Gamma\ \textgreater\ \tau\Gamma^\prime\ and\ \Gamma \le\ \rho \rightarrow USF-MUE, \label{Case1}\\
& \rm If\ \Gamma\ \textgreater\ \tau\Gamma^\prime\ and\ \Gamma\ \textgreater\ \rho \rightarrow CSF-MUE, \\
& \rm If\ \Gamma \le \tau\Gamma^\prime\ and\ \Gamma^\prime\ \textgreater\ \rho^\prime \rightarrow USF-UUE,\\
& \rm If\ \Gamma \le \tau\Gamma^\prime\ and\ \Gamma^\prime \le\ \rho^\prime \rightarrow CSF-UUE.\label{Case4}
\end{align}
%

Once a UE is assigned to a MOI/UOI and is scheduled within the USF/CSF radio frames, then the SE for this UE can be expressed for the four different scenarios in \eqref{Case1}-\eqref{Case4} as follows:
\begin{align}
C_{\rm usf}^{\rm mbs} &= \frac{\beta {\rm log_2}(1+\Gamma)}{N_{\rm usf}^{\rm mbs}},~\label{Cap1}\\
C_{\rm csf}^{\rm mbs} &= \frac{(1-\beta){\rm log_2}(1+\Gamma_{\rm csf})}{N_{\rm csf}^{\rm mbs}},\\
C_{\rm usf}^{\rm uabs} &= \frac{\beta {\rm log_2}(1+\Gamma^\prime)}{N^{\rm uabs}_{\rm usf}},~\\
C_{\rm csf}^{\rm uabs} &= \frac{(1-\beta){\rm log_2}(1+\Gamma^\prime_{\rm csf})}{N^{\rm uabs}_{\rm csf}},\label{Cap4}
\end{align}
where $N_{\rm usf}^{\rm mbs}$, $N_{\rm csf}^{\rm mbs}$, $N^{\rm uabs}_{\rm usf}$, and $N^{\rm uabs}_{\rm csf}$ are the number of MUEs and UUEs scheduled in USF and CSF radio subframes, and $\Gamma$, $\Gamma_{\rm csf}$, $\Gamma^\prime$, $\Gamma^\prime_{\rm csf}$ are as in \eqref{Eq:SIR1}-\eqref{Eq:SIR4}. 

In this paper, we consider the use of 5pSE which corresponds to the worst fifth percentile UE capacity among the capacities of all the $N_{\rm ue}$ UEs (calculated based on \eqref{Cap1}-\eqref{Cap4}) within the simulation area. We believe it is a critical metric particularly for PSC scenarios to maintain a minimum QoS level at all the UEs in the environment.
We define the dependency of the 5pSE to UABS locations and ICIC parameters as 
\begin{align}
C_{\rm 5th}\Big({\bf X}_{\rm uabs},{\bf S}_{\rm mbs}^{\rm ICIC},{\bf S}_{\rm uabs}^{\rm ICIC}\Big)~,
\end{align}
where ${\bf X}_{\rm uabs}\in \mathbb{R}^{N_{\rm uabs}\times 3}$ captures the UABS locations as defined earlier, ${\bf S}_{\rm mbs}^{\rm ICIC} = [\boldsymbol{\alpha},\boldsymbol{\rho}]$ $\in \mathbb{R}^{N_{\rm mbs} \times 2}$ is a matrix that captures individual ICIC parameters for each MBS, while ${\bf S}_{\rm uabs}^{\rm ICIC} = [\boldsymbol{\tau},\boldsymbol{\rho'}]$ $\in \mathbb{R}^{N_{\rm uabs} \times 2}$ is a matrix that captures individual ICIC parameters for each UABS. In particular,  
\begin{align}
\boldsymbol{\alpha}=[\alpha_1,...,\alpha_{N_{\rm mbs}}]^T,\quad \boldsymbol{\rho}=[\rho_1,...,\rho_{N_{\rm mbs}}]^T
\end{align}
are $N_{\rm mbs}\times 1$ vectors that include the power reduction factor and MUE scheduling threshold parameters for each MBS. On the other hand, 
\begin{align}
\boldsymbol{\tau}=[\tau_1,...,\tau_{N_{\rm uabs}}]^T, \quad \boldsymbol{\rho}^\prime=[\rho_1^\prime,...,\rho_{N_{\rm uabs}}^\prime]^T
\end{align}
are $N_{\rm uabs}\times 1$ vectors that involve the CRE bias and UUE scheduling threshold at each UABS. 
As noted in Section~\ref{icicidetails}, the duty cycle $\beta$ of ABS and reduced power subframes is assumed to be set to $0.5$ at all MBSs to reduce search space and complexity. 

Considering that the optimum values of the vectors $\boldsymbol{\alpha}$, $\boldsymbol{\rho}$, $\boldsymbol{\rho}'$, and $\boldsymbol{\tau}$ are to be searched over a multi-dimensional space, computational complexity of finding the optimum parameters is prohibitively high. Hence, to reduce the system complexity (and simulation runtime) significantly, we consider the same ICIC parameters are used for all MBSs and for all UABSs. In particular, we consider that for $i=1,...,N_{\rm mbs}$ we have~$\alpha_i=\alpha$ and $\rho_i=\rho$, while for $j=1,...,N_{\rm uabs}$ we have $\tau_j=\tau$ and $\rho_j^\prime=\rho^\prime$. Therefore, the dependence of the 5pSE on the UABS locations and ICIC parameters can be simplified as
\begin{align}
C_{\rm 5th}\big({\bf X}_{\rm uabs},\alpha,\rho,\tau,\rho^\prime\big)~,
\end{align}
which we will seek ways to maximize in the next section. We leave the problem of individually optimizing ICIC parameters for the MBSs and UABSs as a future work due to the high computational complexity of the problem. 
\begin{figure} [t]
\centering
\includegraphics[width=0.85\linewidth]{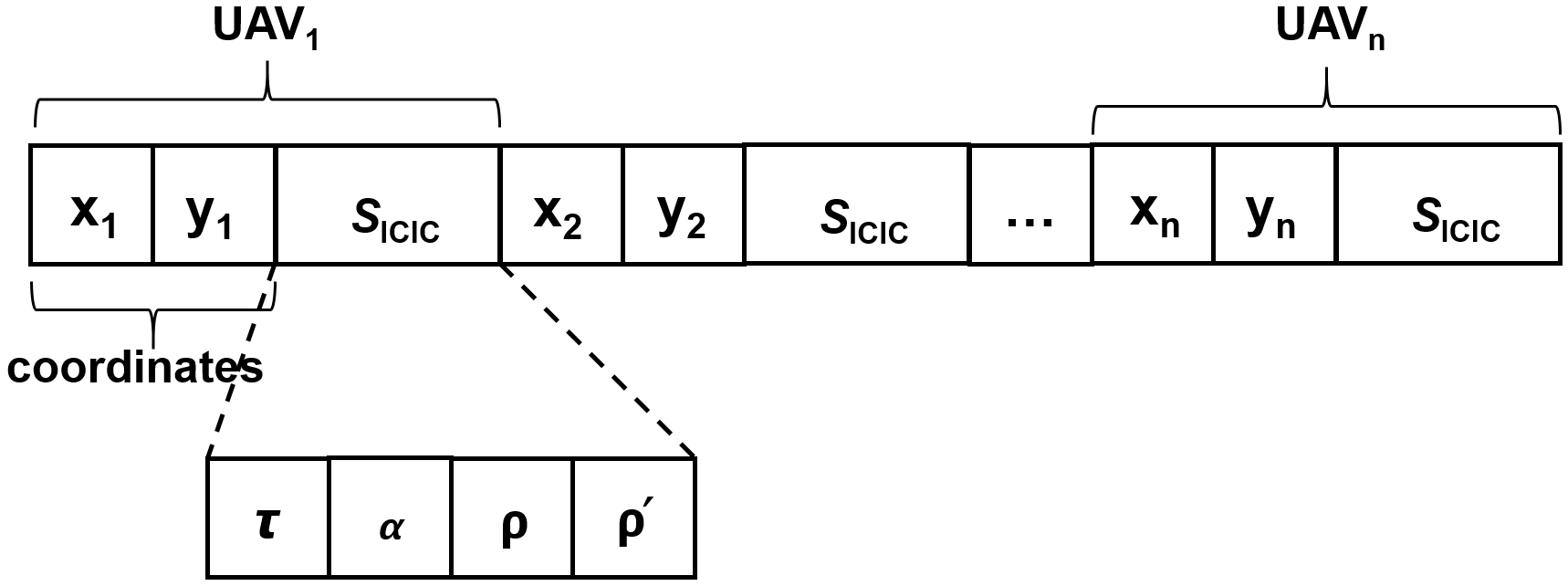}
\caption{An example of a chromosome for FeICIC simulation, where the UABS locations, ICIC parameter $\tau$, $\alpha$, $\rho$, and $\rho^\prime$ are optimized. The ICIC parameter $\beta$ is not optimized and is fixed at 50\% duty cycle.}
\label{GAChromosome} \vspace{-0.2cm}
\end{figure}

\begin{lstlisting}[caption=Steps for optimizing population using GA., basicstyle=\scriptsize, language=R, breaklines=true, numbers=none, frame=single, showstringspaces=false, xleftmargin=0.2cm, linewidth=8.7cm,label=GaListing]
Input:
Population: set of UABS locations and ICIC parameters
FITNESS function: 5pSE of the network
Output:
Args: Best individuals of ICIC parameters and
      highest 5pSE
Method:
NewPopulation <- empty set
StopCondition: Number of iterations = 6
SELECTION: Roulette wheel selection method
while(! StopCondition)
{
  for i = 1 to Size do
  {
    Parent1 <- SELECTION(NewPopulation,FITNESS function)
    Parent2 <- SELECTION(NewPopulation,FITNESS function)
    Child <- Reproduce(Parent1, Parent2)
    if(small random probability)
    {
     child <- MUTATE(Child)
     add child to NewPopulation
    }
  }
  EVALUATE(NewPopulation, FITNESS function);
  Args <- GetBestSolution(NewPopulation)
  Population <- Replace(Population, NewPopulation)
}
\end{lstlisting}

\section{UABS Deployment Optimization}
\label{sec:UabsDeploy}
In this section, we discuss the UABS deployment using the genetic algorithm (GA) and the hexagonal grid model, where we use the 5pSE as an optimization metric to maximize for both scenarios.

\subsection{Genetic Algorithm based UABS Deployment Optimization}
\label{GaSection}

The GA is a population-based optimization technique that can search a large environment simultaneously to reach an optimal solution~\cite{merwaday2016improved}. In this paper, the UABS coordinates and the ICIC parameters constitute the GA population, and a subsequent chromosome is illustrated in Fig.~\ref{GAChromosome}. Applying the detailed steps described in~\cite{merwaday2016improved}, Listing~\ref{GaListing} describes the main steps used to optimize the UABS locations and ICIC parameters while computing the 5pSE.

We apply the GA to simultaneously optimize the UABS locations and ICIC network parameters in order to maximize the 5pSE of the network over a given geographical area of interest. The location of each UABS within a rectangular simulation area is given by $(x_i,y_i)$ where $i \in \{ 1,2,...,N_{\rm uabs}\}$. The UABS locations and the ICIC parameters that maximize the 5pSE objective function can be calculated as
%
\begin{align}
\big[\hat{\bf X}_{\rm uabs},\hat{\alpha},\hat{\rho},&\hat{\tau},\hat{\rho^\prime}\big]= \nonumber\\
&\arg\underset{{\bf X}_{\rm uabs},\alpha,\rho,\tau,\rho^\prime}{\max} C_{\rm 5th}\big({\bf X}_{\rm uabs},\alpha,\rho,\tau,\rho^\prime\big). \label{GA_Optim}
\end{align}
Since searching for optimal ${\bf X}_{\rm uabs}$ and ICIC parameters through a brute force approach is computationally intensive, in this paper we use the GA to find optimum UABS locations and the best-fit ICIC parameters $\tau$, $\alpha$, $\rho$, and $\rho^\prime$.


\begin{lstlisting}[caption=Steps for computing 5pSE for hexagonal grid deployment., basicstyle=\scriptsize, language=R, breaklines=true, numbers=none, frame=single, showstringspaces=false, xleftmargin=0.2cm, linewidth=8.7cm, label=HexListing]
Input: set of UABS locations and ICIC parameters
Output: SE: 5pSE for the network
Method:
StopCondition: Number of iterations = 100
while(! StopCondition)
{
  Generate UABS locations
  for t = 1 to ICICParms.tau[t] do
  {
    for a = 1 to ICICParms.alpha[a] do
    {
      for r = 1 to ICICParms.rho[r] do
      {
        for p = 1 to ICICParms.rhoprime[p] do
        {
          SE = Calc5thPercentileSE(nodal locations, nodal Tx powers, path-loss, tau, beta, alpha, rho, rhoprime)
        }
      }
    }
  }
}
\end{lstlisting}

\begin{figure}[h]
\centering
\includegraphics[width=0.4\textwidth]{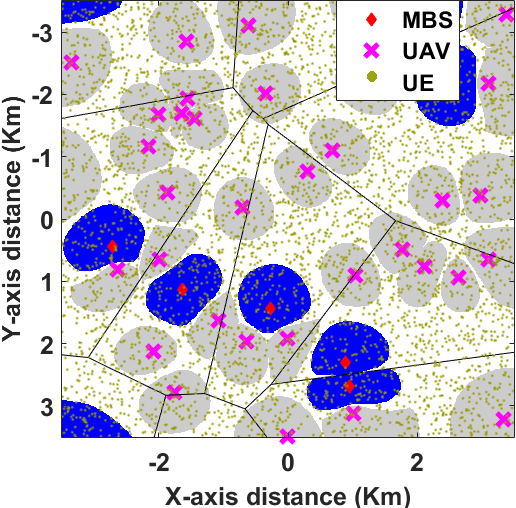}
\caption{PSC network after a disaster with UABS locations optimized using the GA (UAV height: 100~m).}
\label{fig:infrastructure_GA}
\end{figure}

For the MBS locations shown in Fig.~\ref{fig:infrastructure1}(b), an example outcome of UABS locations using the GA is shown in  Fig.~\ref{fig:infrastructure_GA}. Given the mobility and agility of UABSs, using the GA, the UAV positions can be dynamically rearranged to optimized locations to achieve the best network performance at the site of an emergency. 

\subsection{UABS Deployment in a Hexagonal Grid}
\label{HexSection}
As a lower complexity alternative to optimizing UABS locations, we consider deploying the UABSs on a hexagonal grid, where the positions of UABSs are deterministic. We assume that the UABSs are placed within the rectangular simulation area regardless of the existing MBS locations. The 5pSE for this network is determined by using a brute force technique as described in the Listing~\ref{HexListing} which only considers optimization of the ICIC parameters captured through the matrix ${\bf S}_{\rm ICIC}$. In particular, the ICIC parameters that maximize the 5pSE can then be calculated as:
\begin{equation}
\big[\hat{\alpha},\hat{\rho},\hat{\tau},\hat{\rho^\prime}\big]=\arg\underset{\alpha,\rho,\tau,\rho^\prime}{\max} ~ C_{\rm 5th}\big({\bf X}^{\rm (hex)}_{\rm uabs},\alpha,\rho,\tau,\rho^\prime\big),\label{Hex_Optim}
\end{equation}
where ${{{\bf X}_{\rm uabs}^{\rm (hex)}}}$ are the fixed and known hexagonal locations of the deployed UABSs within the simulation area. 

\begin{figure}
\centering
\includegraphics[width=0.4\textwidth]{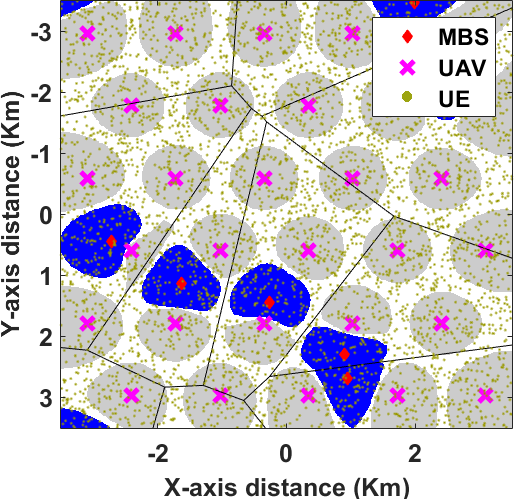}
\caption{PSC network after a disaster with UABS deployed on a fixed hexagonal grid (UAV height: 100~m).}
\label{fig:infrastructure_Hex}
\end{figure}

For the  MBS locations shown in Fig.~\ref{fig:infrastructure1}(b), an example outcome of the UABS locations using the hexagonal grid deployment is shown in  Fig.~\ref{fig:infrastructure_Hex}. In the case of loss in PSC network infrastructure, the UABSs can be deployed rapidly on a fixed hexagonal grid as a primary deployment strategy to form new small cells and consequently
improve the network coverage.

\section{Simulation Results}
\label{simulation}
In this section, using extensive computer simulations, we compare the 5pSE of a UABS-assisted PSC HetNet with and without ICIC techniques while considering different UABS deployment strategies and path loss models for all the UEs covered by the base stations. Unless otherwise specified, the system parameters for the simulations are set to the values given in Table~\ref{tab:SysParams}.

\begin{table}[t]
\caption{Simulation parameters.}
\label{tab:SysParams}
\centering
\footnotesize
\begin{tabular}{|p{4.85cm}|p{3.25cm}|}
\hline
{\bf Parameter} & {\bf Value}  \\ \hline
MBS and UE intensity & $4$ per km$^2$ and $100$ per km$^2$ \\ \hline
MBS and UABS transmit powers & $46 {\rm\ dBm}$ and $30 {\rm\ dBm}$\\ \hline
Path-loss exponent & $4$\\ \hline
Altitude of MBSs & $30 {\rm\ m}$\\ \hline
Altitude of UABSs & $100 {\rm\ m}$\\ \hline
Height of UE & 3 m \\ \hline
PSC LTE Band~14 center frequency & 763 MHz for downlink and 793 MHz for uplink \\ \hline
${\rm d_{mn}^{min}, d_{mu}^{min}}$ & $30 {\rm\ m}$, $10 {\rm\ m}$ \\ \hline
Simulation area & $10 \times 10 {\rm\ km^2}$\\ \hline
GA population size and generation number & $60$ and $100$\\\hline
GA crossover and mutation probabilities & $0.7$ and $0.1$\\\hline
Cell range expansion ($\tau$) in dB & $0$ to $15$ dB \\\hline
Power reduction factor for MBS during ($\alpha$)  &  $0$ to $1$  \\ \hline
Duty cycle for the transmission of USF ($\beta$)  &  $0.5$ or 50\%   \\ \hline
Scheduling threshold for MUEs ($\rho$)            &  $20$ dB to $40$ dB \\ \hline
Scheduling threshold for UUEs ($\rho\prime$)      &  $-20$ dB to $-5$ dB \\ \hline
MBS destroyed sequence & $50\%$ and $97.5\%$ \\\hline
\end{tabular}
\vspace{-2mm}
\end{table}

\subsection{5pSE with UABSs Deployed on a Hexagonal Grid}
\begin{figure*}[t]
\centering
\begin{subfigure}[b]{0.3\textwidth}
\label{HexPppNIM1}
\includegraphics[width=1\textwidth]{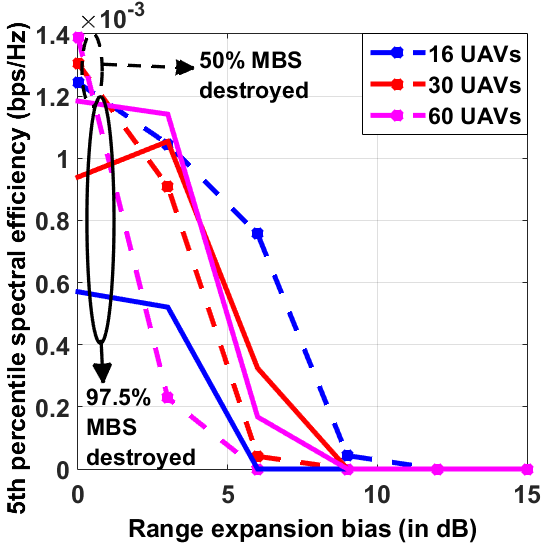}
\caption{5pSE without any ICIC.}
\end{subfigure}
\begin{subfigure}[b]{0.3\textwidth}
\label{HexPppeICIC}
\includegraphics[width=1.02\textwidth]{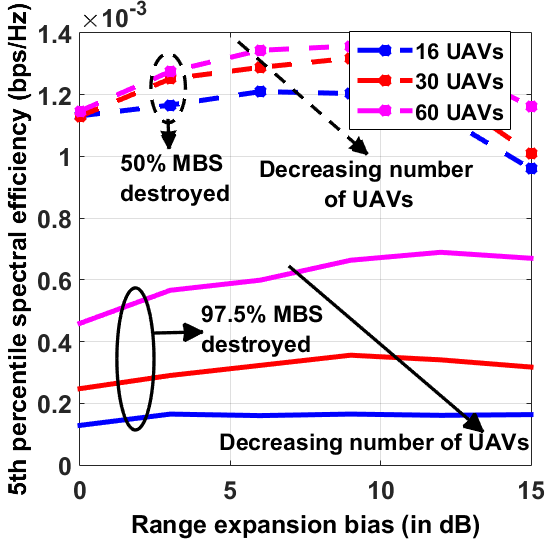}
\caption{5pSE with eICIC.}
\end{subfigure}
\begin{subfigure}[b]{0.3\textwidth}
\label{HexPppFeICIC}
\includegraphics[width=1\textwidth]{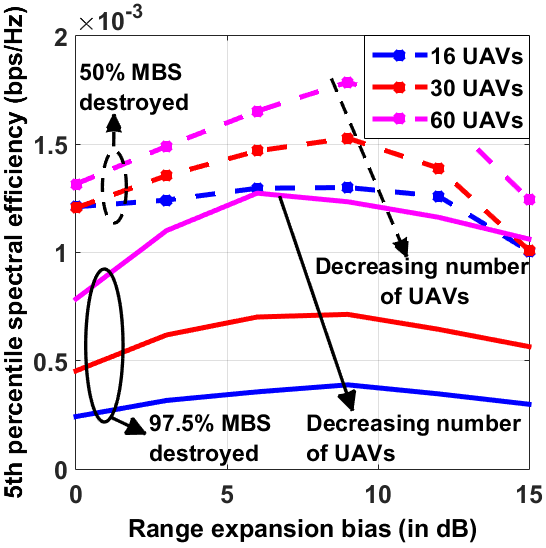}
\caption{5pSE with FeICIC.}
\end{subfigure}
\caption{5pSE versus CRE for eICIC and FeICIC techniques with SPLM (UABSs deployed on a hexagonal grid).}
\label{fig:SE_PPPUE_HexDep}
\end{figure*}

\begin{figure*}[t]
\centering
\begin{subfigure}[b]{0.3\textwidth}
\label{HexPppNIM1}
\includegraphics[width=0.98\textwidth]{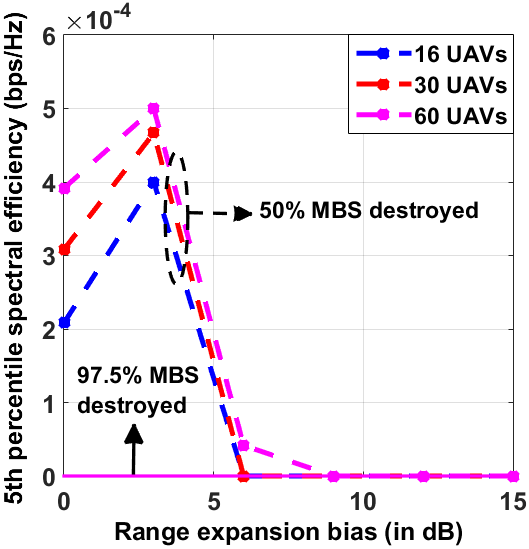}
\caption{5pSE without any ICIC.}
\end{subfigure}
\begin{subfigure}[b]{0.3\textwidth}
\label{HexPppeICIC}
\includegraphics[width=1\textwidth]{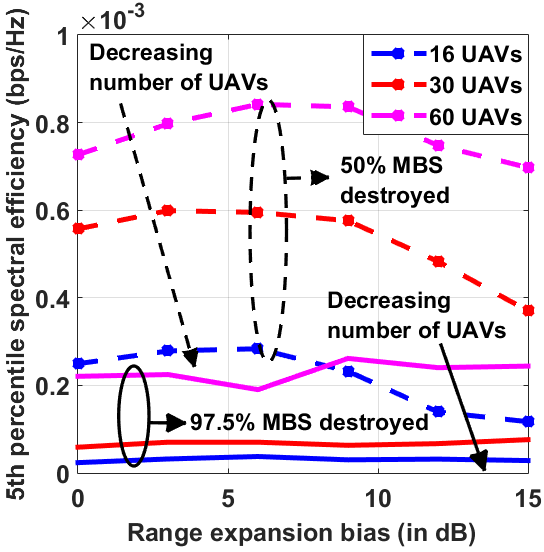}
\caption{5pSE with eICIC.}
\end{subfigure}
\begin{subfigure}[b]{0.3\textwidth}
\label{HexPppFeICIC}
\includegraphics[width=0.97\textwidth]{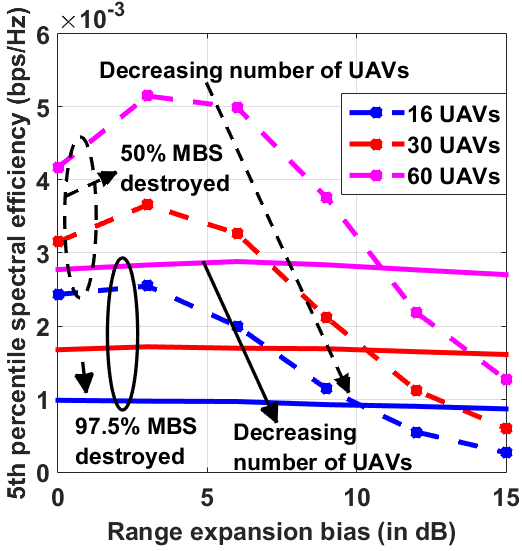}
\caption{5pSE with FeICIC.}
\end{subfigure}
\caption{5pSE versus CRE for eICIC and FeICIC technique with OHPLM (UABSs deployed on a hexagonal grid).}
\label{fig:SE_PPPUE_HexDepOH}
\vspace{-4mm}
\end{figure*}

In the following we will discuss the key 5pSE observations when the UABSs are deployed on a hexagonal grid and utilizing optimized ICIC parameters (see~\eqref{Hex_Optim} and Listing~\ref{HexListing}). In Fig.~\ref{fig:SE_PPPUE_HexDep} and Fig.~\ref{fig:SE_PPPUE_HexDepOH}, we plot the variations in 5pSE with respect to CRE while using SPLM and OHPLM, respectively. 

\begin{figure}[t]
\centering
\includegraphics[width=0.7\linewidth]{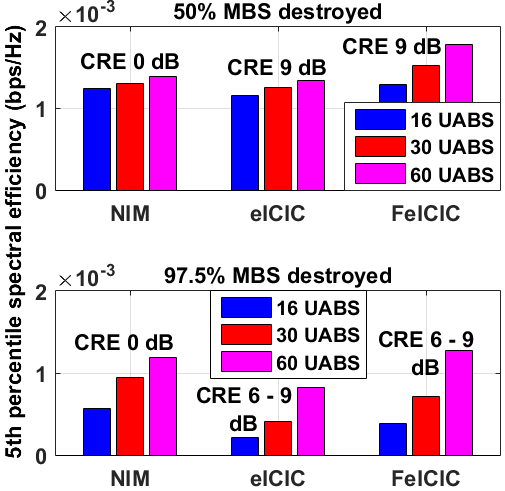}
\caption{Peak observations for the 5pSE with SPLM (UABSs deployed on a hexagonal grid).}
\label{fig:HexGridAna}
\end{figure}

\begin{figure}[h]
\centering
\includegraphics[width=0.7\linewidth]{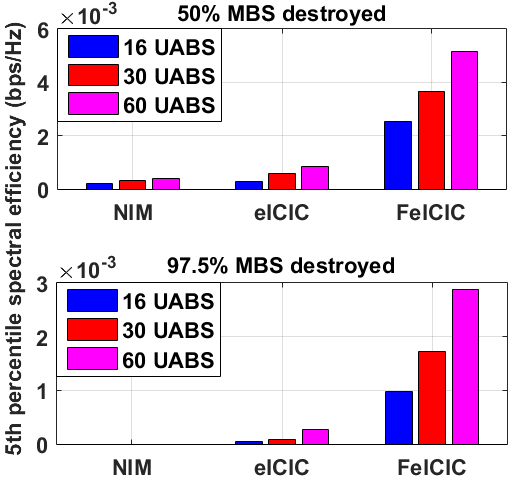}
\caption{Peak observations for the 5pSE with OHPLM (UABSs deployed on a hexagonal grid).}
\label{fig:HexGridAna_OH}
\end{figure}

\subsubsection{5pSE with Simplified Path Loss model}
In Fig.~\ref{fig:SE_PPPUE_HexDep}(a), we plot the 5pSE variation with respect to CRE for no-ICIC mechanism (NIM). In the case of NIM, all the base stations (MBSs and UABSs) always transmit at full power ($P^\prime_{\rm mbs}$ and $P^\prime_{\rm uabs}$). The close evaluation of Fig.~\ref{fig:SE_PPPUE_HexDep}(a), shows that the peak value of 5pSE for NIM is observed at around 0~dB CRE. This is because, with no CRE, the number of UEs associated with the UABSs and the interference experienced by these UEs is minimal. Moreover, as the CRE increases, the number of UEs associated with the UABSs increases and so does the interference experienced by these UEs. Hence, with NIM the 5pSE decreases with increasing CRE as seen in Fig.~\ref{fig:SE_PPPUE_HexDep}(a).

The performance of 3GPP Release-10 and Release-11 ICIC techniques in terms of 5pSE and the variation in CRE are plotted in Fig.~\ref{fig:SE_PPPUE_HexDep}(b) and Fig.~\ref{fig:SE_PPPUE_HexDep}(c). As noted in Section~\ref{icicidetails}, the transmission power during blank subframes at the MBSs for eICIC is $0$, and power reduction of the CSFs at the MBSs for FeICIC is $\alpha P^\prime_{\rm mbs}$. Using this understanding, the analysis Fig.~\ref{fig:SE_PPPUE_HexDep}(b) and Fig.~\ref{fig:SE_PPPUE_HexDep}(c) shows that the 5pSE for ICIC techniques at 0 dB CRE are relatively lower. On the other hand, the ICIC techniques observe improvement in 5pSE performance with increasing CRE and the peak values of the 5pSE for the ICIC techniques is observed when the CRE is between $6-9$~dB. This influence of CRE on the 5pSE for NIM and 3GPP Release-10 and Release-11 ICIC techniques are summarized in Fig.~\ref{fig:HexGridAna}.

\subsubsection{5pSE with Okumura-Hata Path Loss Model}
In Fig.~\ref{fig:SE_PPPUE_HexDepOH}(a), we plot the 5pSE variation with respect to CRE for NIM. In the case of NIM, all the base stations (MBSs and UABSs) always transmit at full power ($P^\prime_{\rm mbs}$ and $P^\prime_{\rm uabs}$). The peak value of 5pSE for NIM is observed at around 3~dB CRE when 50\% of the MBSs are destroyed. On the other hand, when 97.5\% of the MBSs are destroyed, even though the number of existing MBSs are small and the interference is minimum, the higher path-loss presents higher probability for a cell-edge UE to fall out of coverage area. Moreover, in the absence of any ICIC, using CRE can magnify the impact of interference. Hence, the 5pSE gains with NIM are close to zero.

In Fig.~\ref{fig:SE_PPPUE_HexDepOH}(b) and Fig.~\ref{fig:SE_PPPUE_HexDepOH}(c), we plot the variation of 5pSE with respect to CRE for 3GPP Release-10/11 ICIC techniques. The analysis of Fig.~\ref{fig:SE_PPPUE_HexDepOH}(b) and Fig.~\ref{fig:SE_PPPUE_HexDepOH}(c) shows that the ICIC techniques observe improvement in 5pSE performance with increasing CRE. When 50\% MBS are destroyed, the 5pSE peak values for the eICIC and FeICIC are observed when the CRE is between $6-9$~dB and $3-6$~dB, respectively. When 97.5\% of the MBSs are destroyed, even though the cell-edge UEs observe higher path loss, using the 3GPP Release-10/11 ICIC techniques along with CRE can decrease the probability of cell-edge UE going out of coverage. Thus sustaining the 5pSE of the network as seen in Fig.~\ref{fig:SE_PPPUE_HexDepOH}(b) and Fig.~\ref{fig:SE_PPPUE_HexDepOH}(c). Further analysis show that the FeICIC technique observes significant improvement in SE performance when compared to NIM and eICdeIC. This influence of CRE on the 5pSE for NIM and 3GPP Release-10/11 ICIC techniques is summarized in Fig.~\ref{fig:HexGridAna_OH}.

On comparison of Fig.~\ref{fig:HexGridAna} and Fig.~\ref{fig:HexGridAna_OH}, we observe modest deviation in peak values of 5pSE between NIM, eICIC, and FeICIC with SPLM. This is because UEs experience better SIR with lower path-losses. Whereas, with OHPLM we observe significant deviation in the peak values of 5pSE due to higher path-losses. However, the higher path-losses in OHPLM can be compensated by using modest/higher CRE values and 3GPP Release-10/11 ICIC techniques.

Overall, with hexagonal grid deployment for both the path-loss models, the 5pSE for the network is higher when larger number of UABSs are deployed and when fewer MBSs are destroyed. Also, the 5pSE decreases with the increasing number of destroyed MBSs as seen in Fig.~\ref{fig:SE_PPPUE_HexDep} and Fig.~\ref{fig:SE_PPPUE_HexDepOH}.

\begin{figure}
\vspace{-0.2cm}
\centering
\begin{subfigure}[b]{0.24\textwidth}
\label{HexPppeICIC}
\includegraphics[width=1\textwidth]{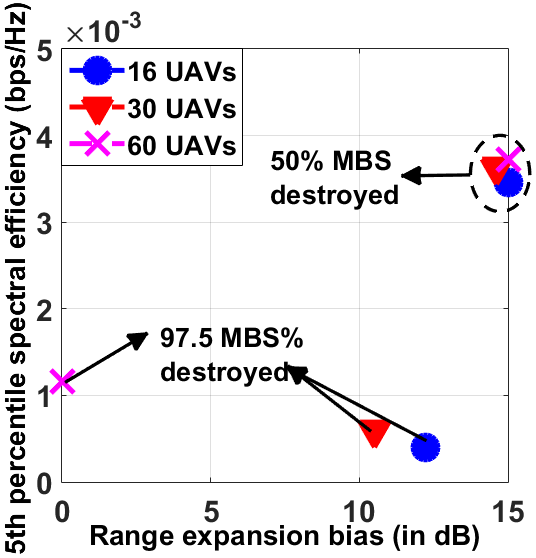}
\caption{5pSE with eICIC.}
\end{subfigure}
\begin{subfigure}[b]{0.24\textwidth}
\label{HexPppFeICIC}
\includegraphics[width=0.99\textwidth]{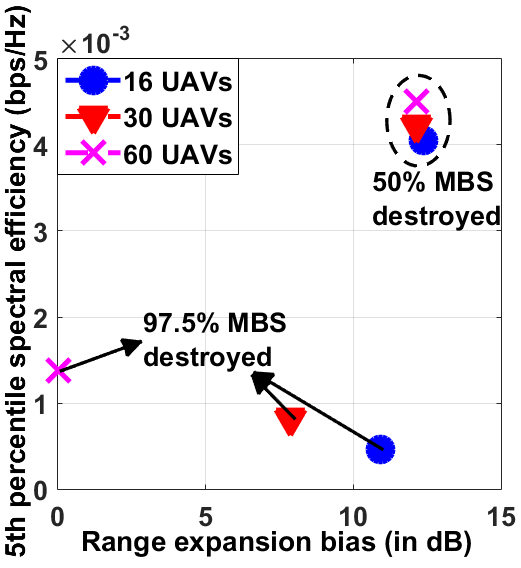}
\caption{5pSE with FeICIC.}
\end{subfigure}
\caption{Peak 5pSE versus optimized CRE for eICIC and FeICIC techniques with SPLM, when the UABS locations and ICIC parameters are optimized using the GA.}
\label{fig:SE_PPPUE_GaDep} \vspace{-0.2cm}
\end{figure}

\begin{figure}
\centering
\begin{subfigure}[b]{0.24\textwidth}
\label{HexPppeICIC}
\includegraphics[width=1\textwidth]{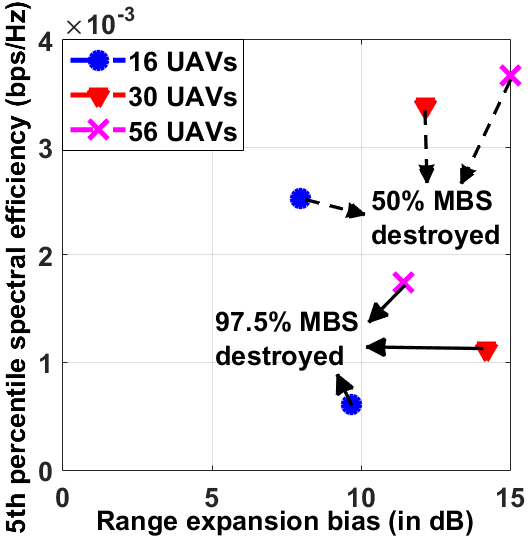}
\caption{5pSE with eICIC.}
\end{subfigure}
\begin{subfigure}[b]{0.24 \textwidth}
\label{HexPppFeICIC}
\includegraphics[width=1.025\textwidth]{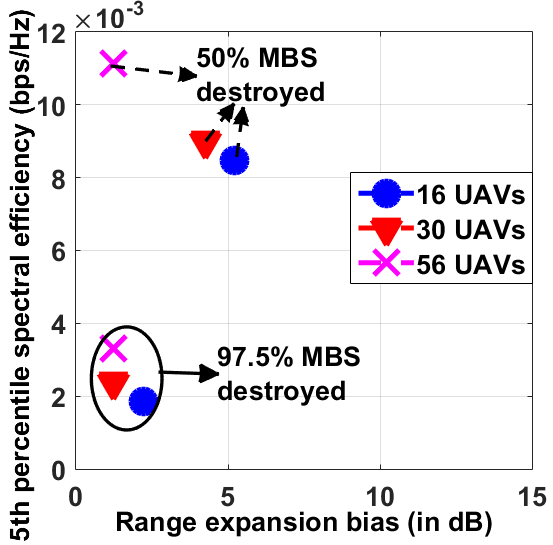}
\caption{5pSE with FeICIC.}
\end{subfigure}
\caption{Peak 5pSE versus optimized CRE for eICIC and FeICIC techniques with OHPLM, when the UABS locations and ICIC parameters are optimized using the GA.}
\label{fig:SE_PPPUE_GaDep_OH} \vspace{-0.2cm}
\end{figure}

\begin{figure} [!htbp]
\centering
\begin{subfigure}[b]{0.24\textwidth}
\label{HexPppeICIC}
\includegraphics[width=1\textwidth]{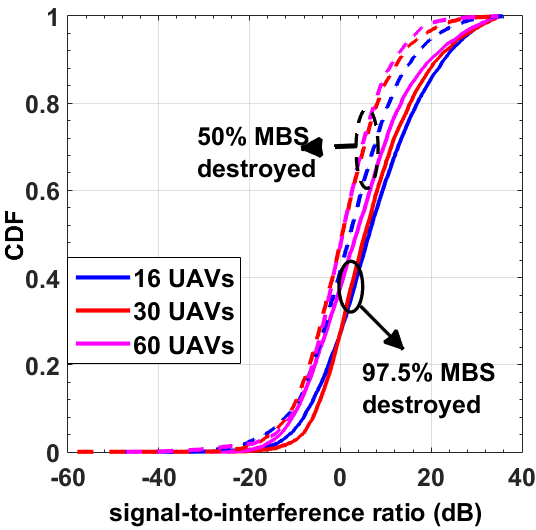}
\caption{SIR observations for eICIC.}
\end{subfigure}
\begin{subfigure}[b]{0.24\textwidth}
\label{HexPppFeICIC}
\includegraphics[width=1\textwidth]{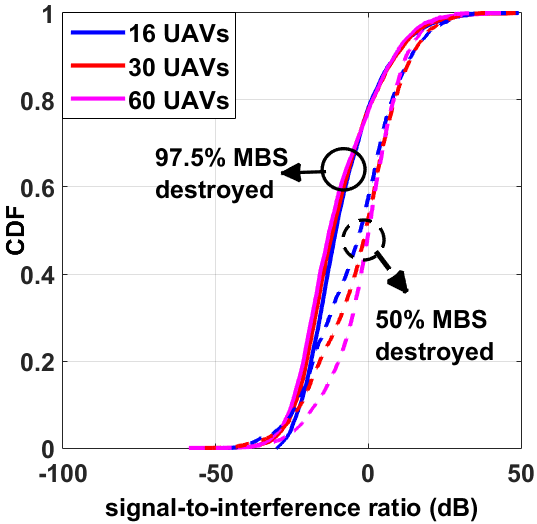}
\caption{SIR observations FeICIC.}
\end{subfigure}
\caption{SIR observations for eICIC and FeICIC with OHPLM, when UABS locations are optimized using the GA.}
\label{fig:OHGA_SIR} \vspace{-0.2cm}
\end{figure}

\subsection{5pSE with GA Based UABS Deployment Optimization}
In the following, we will discuss the key 5pSE observations, when UABS locations and ICIC parameters optimized through the GA as in~\eqref{GA_Optim} and Listing~\ref{GaListing}. In Fig.~\ref{fig:SE_PPPUE_GaDep} and Fig.~\ref{fig:SE_PPPUE_GaDep_OH}, we plot the peak 5pSE for the network using the GA, versus the optimized CRE value while using SPLM and OHPLM, respectively. In the GA based simulations, the optimum CRE value is directly related to the locations of the UABSs with respect to the MBSs, the number of UEs offloaded to the UABSs, and the amount of interference observed by the UEs.

\subsubsection{5pSE with Simplified Path Loss Model} 
In Fig.~\ref{fig:SE_PPPUE_GaDep}(a) and Fig.~\ref{fig:SE_PPPUE_GaDep}(b), we plot the peak 5pSE with respect to the optimized CRE value for eICIC and FeICIC, respectively, for SPLM. Inspection of Fig.~\ref{fig:SE_PPPUE_GaDep}(a) and Fig.~\ref{fig:SE_PPPUE_GaDep}(b) shows higher values of CRE when $50\%$ of the MBSs are destroyed and implies the presence of substantial interference from these large number of MBSs. Hence, offloading a large number of UEs from MBSs to UABSs with higher values of CRE is necessary for achieving better 5pSE gains.

On the other hand, when most of the infrastructure is destroyed (i.e., when $97.5\%$ of the MBSs destroyed), the interference observed from the MBSs is limited, and a larger number of UEs need to be served by the UABSs. Therefore, with fewer UABSs deployed, higher CRE is required to serve a larger number of UEs and achieve better 5pSE. On the other hand, when a larger number of UABSs are deployed, smaller values of CRE will result in better 5pSE gains. We record these behavior in Fig.~\ref{fig:SE_PPPUE_GaDep}(a) and Fig.~\ref{fig:SE_PPPUE_GaDep}(b) for eICIC and FeICIC, respectively.

\begin{figure}
\begin{center}
\begin{subfigure}[b]{0.42\textwidth}
\label{HexPppeICIC}
\begin{center}
\includegraphics[width=0.8\textwidth]{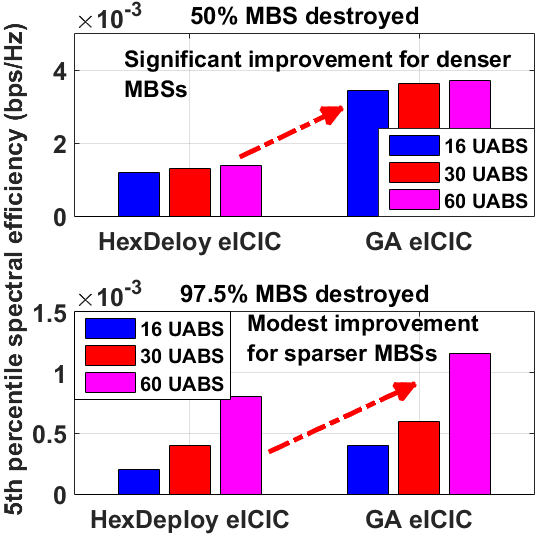}
\caption{5pSE with eICIC.}\vspace{3mm}
\end{center}
\end{subfigure}
\begin{subfigure}[b]{0.42\textwidth}
\label{HexPppFeICIC}
\begin{center}
\includegraphics[width=0.8\textwidth]{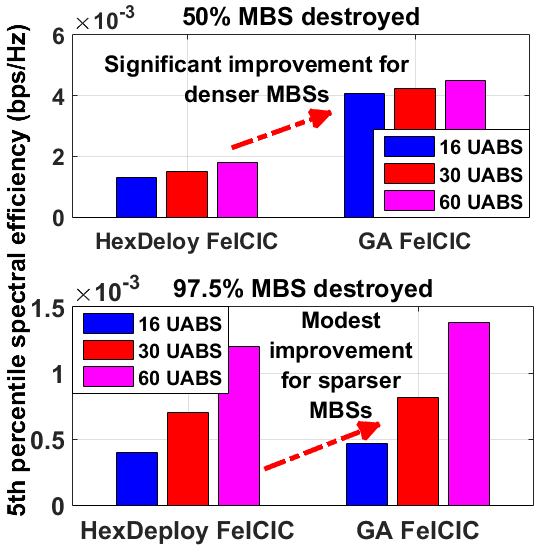}
\caption{5pSE with FeICIC.}
\end{center}
\end{subfigure}
\caption{5pSE comparisons for eICIC and FeICIC with SPLM, when the UABS locations are optimized using the GA and when the UABSs are deployed in a fixed hexagonal grid.}
\label{fig:SE_HexVsGa}
\end{center}
\end{figure}

\begin{figure}
\begin{center}
\begin{subfigure}[b]{0.42\textwidth}
\label{HexPppeICIC}
\begin{center}
\includegraphics[width=0.8\textwidth]{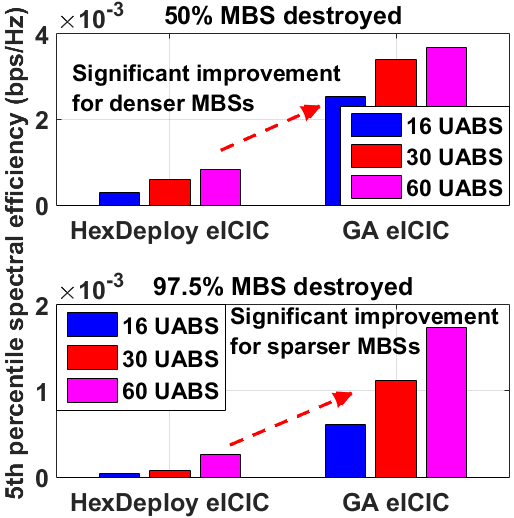}
\caption{5pSE with eICIC.}\vspace{3mm}
\end{center}
\end{subfigure}
\begin{subfigure}[b]{0.42\textwidth}
\label{HexPppFeICIC}
\begin{center}
\includegraphics[width=0.8\textwidth]{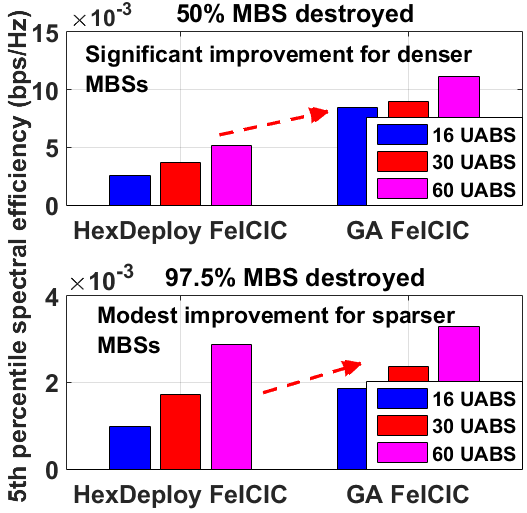}
\caption{5pSE with FeICIC.}
\end{center}
\end{subfigure}
\caption{5pSE comparisons for eICIC and FeICIC with OHPLM, when the UABS locations are optimized using the GA and when the UABSs are deployed in a fixed hexagonal grid.}
\label{fig:SE_HexVsGa_OH}
\end{center}
\end{figure}

\subsubsection{5pSE with Okumura-Hata Path Loss model}
Using \eqref{Eq:SIR1}--\eqref{Eq:SIR4}, we plot the SIR observations in Fig.~\ref{fig:OHGA_SIR} for 3GPP Release-10/11 ICIC techniques. As illustrated in Fig.~\ref{fig:pathloss}(a), the higher path-loss results in lower SIR values as seen in Fig.~\ref{fig:OHGA_SIR}. Using this understanding, we inspect the peak 5pSE with respect to the optimized CRE for the eICIC and FeICIC as shown in Fig.~\ref{fig:SE_PPPUE_GaDep_OH}(a) and Fig.~\ref{fig:SE_PPPUE_GaDep_OH}(b), respectively.

With 3GPP Release-10 ABS, higher values of CRE are required for UABSs to compensate for the high path-loss and under-utilization of radio resources by the MBSs in CSF radio subframes. When $50\%$ and $97.5\%$ of the MBSs are destroyed, the peak 5pSE values for eICIC are achieved with minimal SIR values, and by offloading a large number of UEs from MBSs to UABSs as seen in Fig.~\ref{fig:SE_PPPUE_GaDep_OH}(a).

On the other hand, with 3GPP Release-11 reduced power subframes (FeICIC), MBSs can establish and maintain connectivity with sufficient number of cell-edge MUEs, while offloading the out-of-coverage UEs to UABSs for better QoS. When $50\%$ and $97.5\%$ of the MBSs are destroyed, the peak 5pSE values for FeICIC are achieved with minimal SIR and moderate CRE values as shown in Fig.~\ref{fig:SE_PPPUE_GaDep_OH}(b).

To summarize, using GA for both path-loss models, FeICIC in Release-11 is seen to outperform Release-10 eICIC in  terms of the overall 5pSE of the network. When larger number of UABSs are deployed and when fewer MBSs are destroyed, 5pSE of the network is higher. On the other hand, the 5pSE decreases with the increasing number of destroyed MBSs as seen in~Fig.~\ref{fig:SE_PPPUE_GaDep_OH}.

\subsection{Performance Comparison Between Fixed (Hexagonal) and Optimized UABS Deployment with eICIC and FeICIC}
We summarize our key results from earlier simulations in Fig.~\ref{fig:SE_HexVsGa} and Fig.~\ref{fig:SE_HexVsGa_OH} for both path-loss models and compare the key trade-offs between fixed (hexagonal) deployment and GA based deployment of UABSs. 

\subsubsection{Influence of SPLM  on 5pSE}
In Fig.~\ref{fig:SE_HexVsGa}, we compare the 5pSE observations with SPLM shown in Fig.~\ref{fig:SE_PPPUE_HexDep} and Fig.~\ref{fig:SE_PPPUE_GaDep}. The comparative analysis reveals that UABSs deployment with optimized CRE and optimized location provides a better 5pSE than the UABSs deployed on a fixed hexagonal grid. Furthermore, Fig.~\ref{fig:SE_HexVsGa} shows that the 5pSE gains from the optimization of UABS locations are more significant when $50\%$ of the MBSs are destroyed and less significant when $97.5\%$ of the MBSs are destroyed. 

When $50\%$ MBSs are destroyed, there are still a large number of MBSs present which causes substantial interference. Hence, in such interference driven scenario it is important to optimize the locations of the UABSs, and use of larger number of UABSs to provide significant gains in the 5pSE. On the other hand, with $97.5\%$ of the MBSs destroyed, the interference from the MBSs is small, and deploying the UABSs on a hexagonal grid will perform close to optimum UABS deployment. 

\subsubsection{Influence of OHPLM on 5pSE}
In Fig.~\ref{fig:SE_HexVsGa_OH}, we compare the 5pSE observations with OHPLM shown in Fig.~\ref{fig:SE_PPPUE_HexDepOH} and Fig.~\ref{fig:SE_PPPUE_GaDep_OH}. The comparative analysis reveals that UABSs deployment with optimized CRE and optimized location provides a better 5pSE than the UABSs that are deployed on a fixed hexagonal grid. 

With eICIC in Release-10, when $50\%$ and $97.5\%$ of the MBSs are destroyed, the 5pSE gains from the optimized UABS locations are significant as shown in Fig.~\ref{fig:SE_HexVsGa_OH}(a). On the other hand, with FeICIC in Release-11, the 5pSE gains from the optimized UABS locations are more significant when $50\%$ of the MBSs are destroyed as seen in Fig.~\ref{fig:SE_HexVsGa_OH}(b). However, the difference between the hexagonal deployment and optimized deployment is especially small since the power reduction factor $\alpha$ in the MBS CSFs provides an additional optimization dimension for improving the 5pSE. Use of a larger number of UABSs when $97.5\%$ of the MBSs are destroyed is shown to provide modest gains in the 5pSE, in contrast to significant gains in the 5pSE when $50\%$ of the MBSs are destroyed.

\begin{figure} [t]
\centering
\includegraphics[width=0.8\linewidth]{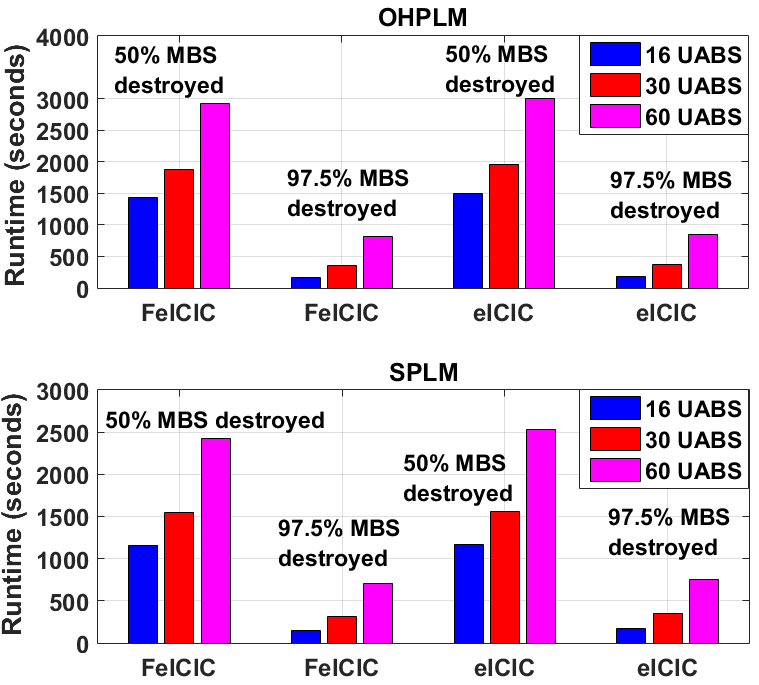}
\caption{GA simulation runtime using FeICIC and eICIC technique with OHPLM and SPLM.}
\label{GARuntime} \vspace{-0.2cm}
\end{figure}

\begin{figure} [t]
\centering
\includegraphics[width=0.83\linewidth]{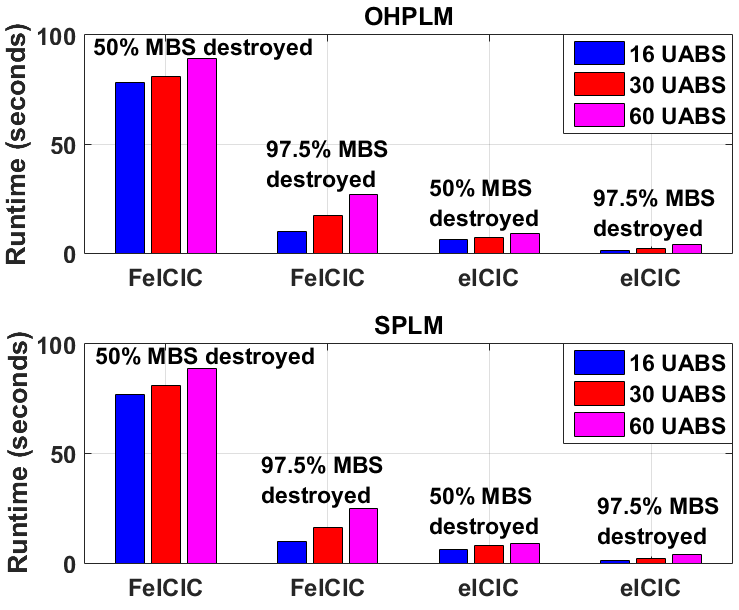}
\caption{Fixed hexagonal grid simulation runtime using FeICIC and eICIC technique with OHPLM and SPLM.}
\label{HexRuntime} \vspace{-0.2cm}
\end{figure}

\subsection{Comparison of Computation Times for Different UABS Deployment Algorithms}

In this subsection, we compare the computation times for the GA and hexagonal grid deployment techniques with ICIC optimization. Using an Intel Core i7-4810 central processing unit operating at 2.8 GHz, 24 GB of random access memory, and Monte-Carlo experimental approach, we calculate the mean runtime for the Matlab simulations. 
In Fig.~\ref{GARuntime}, we plot the mean runtime required for calculating the optimal ICIC network parameters and optimized UABS locations using ~\eqref{GA_Optim}, Listing~\ref{GaListing}, and the simulation values defined in Table~\ref{tab:SysParams}. Inspection of Fig.~\ref{GARuntime} reveals FeICIC technique requires similar computational time when compared to the eICIC technique for GA based optimization. The main reason for this is that the large search space for UABS locations, which are common in both FeICIC and eICIC based approaches, dominates the computation time when compared with the optimization of ICIC parameters.

On the other hand, in Fig.~\ref{HexRuntime}, we plot the mean runtime required for UABS deployment on a hexagonal grid using~\eqref{Hex_Optim}, the simulation values defined in Table~\ref{tab:SysParams}, and with fixed step size for the ICIC parameters. Inspection of Fig.~\ref{HexRuntime} reveals FeICIC technique requires significantly higher computational time when compared to the eICIC technique. The main reason for this behavior is due to additional computation required for optimizing the power reduction factor $\alpha$ for the FeICIC approach. In general, with the GA and the hexagonal grid deployment, when  larger number of UABSs are deployed, and larger number of the MBSs are present, the mean runtime is the largest. On the other hand, the mean runtime decreases with smaller number of UABSs deployed and when a smaller number of MBSs are present. Moreover, the comparative analysis of Fig.~\ref{GARuntime} and Fig.~\ref{HexRuntime} reveals that optimization (ICIC parameters and UABS locations) using GA requires significantly more computational time when compared to UABSs deployment on a hexagonal grid.

To summarize, the GA is a suitable meta-heuristic technique that relies on bio-inspired approach that uses mutations, crossovers, and selections of chromosomes, for finding optimum or close to optimum solution of a search problem. On the other hand, the computational complexity required to optimize the considered UAV deployment optimization problems in real world using the GA techniques require further investigations.

\section{Concluding remarks}
\label{conclusion}
In this article, we show that the mission-critical communications could be maintained and restored by deploying UABSs in the event of any damage to the public safety infrastructure. Through simulations, we compare and analyze the 5pSE of the network for different path-loss models and different UABS deployment strategies. With SPLM, our analysis shows that deployment of the UABSs on a hexagonal grid is close to optimal when the observed interference is limited. In the presence of substantial interference, the GA approach is more effective for deploying UABSs. On the other hand, with OHPLM, the network observes high path-loss when compared to SPLM. To subdue the effects of high path-loss, the GA approach is shown to be more effective. Our simulation shows that optimizing UABSs locations and ICIC parameters using GA yields significant improvement when compared to the deployment of the UABSs on a hexagonal grid.  

Finally, we observe that the HetNets with reduced power subframes (FeICIC) yield better 5pSE than that with almost blank subframes (eICIC). In a simulated network with SPLM and when 60 UABS locations are optimized using the GA, the FeICIC observes a modest improvement over eICIC: approximately $17\%$ and $15\%$  when $50\%$ and $97.5\%$ of the MBSs are destroyed, respectively. On the other hand, with OHPLM and when 60 UABS locations are optimized using the GA, the FeICIC yields a significant improvement over eICIC: approximately $66\%$ and $51\%$ when $50\%$ and $97.5\%$ of the MBSs are destroyed, respectively.  

\section*{Acknowledgment}
This research was supported in part by NSF under the grants AST-1443999 and CNS-1453678. The authors would like to thank Arvind Merwaday for his
helpful feedback. We also thank the Scientific and Technological Research Council of Turkey for supporting Adem Tuncer's research at Florida International University.

\ifCLASSOPTIONcaptionsoff
  \newpage
\fi



%
\bibliographystyle{IEEEtran}
\bibliography{Citations}

\end{document}